\documentstyle[12pt]{article}

\def\data{\number\day/\ifcase\month\or gennaio \or febbraio \or marzo \or
aprile \or maggio \or giugno \or luglio \or agosto \or settembre
\or ottobre \or novembre \or dicembre \fi/\number\year}
\newcommand{\be}{\begin{equation}}
\newcommand{\ee}{\end{equation}}
\newcommand{\fd}{\rightarrow}

\newcommand{\ba}{\begin{array}}
\newcommand{\ea}{\end{array}}
\newcommand{\bq}{\begin{eqnarray}}
\newcommand{\eq}{\end{eqnarray}}
\newcommand{\bqw}{\begin{eqnarray*}}
\newcommand{\eqw}{\end{eqnarray*}}
\newcommand{\ii}{\infty}
\newcommand{\sr}{\stackrel}
\newcommand{\sct}{\scriptstyle}
\newcommand{\sctsct}{\scriptscriptstyle}
\newcommand{\lm}{\lambda}
\begin{document}

\newtheorem{guess}{Theorem}
%%%%%%%%%%%%%%%%%%%%%%%%%%%%%%%%%%%%%%%%%%%%%%%%%%%%%%%%%%%%%%%%%%%%%%%%%%%%%%

\pagenumbering{arabic}

\title{{\bf A novel hierarchy of integrable lattices.}}

\author{  I. Merola$^a$, O. Ragnisco$^b$ and Tu Gui-Zhang$^c$  }

\maketitle

\begin{flushleft}
$^a$ {\small {\it Dipartimento di Fisica, Universit\`a ``La Sapienza'',
Roma ,ITALY }} \\ $\ \ \ ${\small {\it e.mail: Merola @Roma1.INFN.IT}}\\
$^b$ {\small {\it Dipartimento di Fisica, III Universit\`a di Roma, Roma,
ITALY }}\\$\ \ \ $
{\small {\it and
Istituto Nazionale di Fisica Nucleare, Sezione di Roma}}\\
$^c$ {\small {\it Computing Center of Academia Sinica, Beging 100080,
}} \\$\ \ \ ${\small{\it People's Republic of China}}
\end{flushleft}

\[ \]

\begin{center} {\bf Abstract} \end{center}

{\small
In the framework of the reduction technique for Poisson-Nijenhuis
structures, we derive a new hierarchy of integrable lattice, whose
continuum limit is the AKNS hierarchy.\par
In contrast with other differential-difference versions of the AKNS system,
our hierarchy is endowed with a canonical Poisson structure and, moreover,
it admits a vector generalisation.
We also solve the associated spectral problem and explicity contruct
action-angle variables through the r-matrix approach.}

\section{Introduction}
\label{sec1}

The search for discrete integrable systems got a novel impetus
in recent years: the  quantisation of integrable PDE.s [1 - 4],
 the recent findings on integrable discrete-time systems [5 - 7],
including the ``extreme''  case of cellular automata \cite{8},
and even some results in string theory and $2-D$ quantum gravity (related
to the so-called ``discrete string equations'') are  perhaps the main
motivations for such a growing interest in that field.
The results reported in this paper can be ascribed to the above line of
research, although they belong in a sense to a more traditional
approach, aiming at deriving integrable systems with discrete space and
continuous time.\par
In fact, we present and discuss an integrable differential-difference
version of so-called AKNS hierarchy \cite{9}, already mentioned in
\cite{9bis}.
In comparison with other
discretisations \cite{10} \cite{3},
it has certain advantages and some drawbacks. The latters are  essentially
given by the non-existence of one-field reductions, unlike the model derived
by Ablowitz and Ladik \cite{10}:
 such reductions are in fact admissible only in the continuum
limit.
However, this seems the price to be paid (i) to preserve the first
hamiltonian structure of the AKNS hierarchy, namely the canonical one, and
(ii) to allow for a vector generalisation, both such features being
exhibited by our model.
We have  to mention that one equation belonging to our hierarchy can be
interpreted as a Backlund tranformations for the continuous AKNS
system, and, as such, it has been recently derived by Yamilov
and Svinolupov \cite{11}.
\newline
The paper is organised as follows.
\par
In Section \ref{sec2}, the hierarchy under scrutiny is derived
by using a nowadays standard geometrical reduction
technique for Poisson-Nijenhuis structures.
\par
In Section  \ref{sec3}, the underlying direct and inverse problem
is solved: for the sake of simplicity, only the $2\times 2$ matrix case is
considered.\par
In Section \ref{sec4}, the r-matrix structure of the system is revealed, and,
through the \linebreak  r-matrix, the action-angle variables are explicitly
given in terms of the spectral data.
\par
In section \ref{sec5}, it is shown that the whole hierarchy goes  into
the AKNS one in a suitable simple continuum limit.
\par
In Section \ref{sec6} we make a few concluding remarks and mention some
interesting open problems.

\section{Poisson-Nijenhuis structure}
\label{sec2}

In a recent paper, P.M.Santini and one of the authors \cite{R},
have considered the following abstract linear problem:
\be
E \psi = (Q +\lambda A)\psi
\label{2.1}
\ee
In formula (\ref{2.1}), $\lambda \in \bf{C}$ is a spectral parameter,
$Q$, $A$, $\psi$ take values in an associative algebra ${\cal A}$ with
unit element $I$, endowed with a trace-form $<\cdot,\cdot>$,
$E$ is an algebra automorphism;
$A$ is assumed to be a constant ($\doteq$ fixed point of $E$)
element in the algebra.
\par
It has been shown in \cite{R} that to the linear problem (\ref{2.1}) one
can naturally associate the following two compatible Poisson tensors:
\bq
\theta_{1} &\doteq & L_Q R_A - R_Q L_A \label{2.2a} \\
\theta_2 &\doteq &{\cal N} \theta_1 \label{2.2b}
\eq
where the symbol $L_x$ (resp.$R_x$) denote {\it left} (resp. {\it right})
multiplication by $x$, and ${\cal N}$
is the hereditary recursion operator, or Nijenhuis tensor \cite{N},
defined as:
\be
{\cal N} \doteq
(L_Q E-R_Q) (L_A E - R_A)^{-1}
\label{2.3}
\ee
Furthermore, one has proved there that the family of vector fields:
\be
K_B \doteq (R_B - L_B )Q \doteq[Q,B]; \ \ \ \ \ [A,B]=0
\label{2.4}
\ee
is an invariant family for ${\cal N}$, namely the Lie-derivative
of ${\cal N}$ along $K_B$  vanishes:
\be
{\cal L}_{K_B} {\cal N}= 0
\label{2.5}
\ee
We can then assert that to the linear problem (\ref{2.1}) it is naturally
associated the following hierarchy of (commuting) bi-hamiltonian systems:
\be
{\partial Q \over \partial t_s}= {\cal N}^s K_B
\label{2.6}
\ee
Out of the abstract hierarchy (\ref{2.6}), one can construct bi-hamiltonian
lattices through a convenient realization of the algebra \( {\cal A} \),
chosen to be the algebra of matrix valued sequences, approaching an arbitrary
constant value as $|n| \rightarrow \infty $, and of the automorphism $E$,
taken to be the ``shift operator'' on sequences:
\be
E X_n=X_{n+1}\ \ \ \ \ \ \ \ \ n\in Z
\label{2.7}
\ee
In particular, throughout this paper, we shall consider $(N+1) \times (N+1)$
real matrices.
Accordingly, the ``field variable'' $Q$ will be parametrized as follows:
\be
Q= \left( \ba{cc} p & <q| \\ |r> & \hat{Q} \ea \right)
\label{2.8}
\ee
where $p$ is a scalar, $<q|$ is the $1\times N$ matrix
$(q_1,.....,q_N)$ (``row vector''), $|r>$ is the  $N\times 1$ matrix
$(r_1,.....,r_N)$ (``column vector'') and $\hat Q$ is an $N\times N$ matrix.
\par
The constant matrix $A$ will be chosen as:
\be
A= \left( \ba{cc} 1 & <0| \\ |0> & \hat{0} \ea \right)
\label{2.9}
\ee
where $<0|$ ($|0>$) is the null row (column) vector, and $\hat{0}$ is the
null $N\times N$matrix.
\par
The sequence $\{Q_n\}_{n\in \bf {Z}} $
 of matrices of type (\ref{2.8}) is assumed to fulfil the boundary
condition:
\be
\lim_{|n| \rightarrow \infty} Q_n = \left( \ba{cc} 0 & <0| \\ |0> & \hat{I}
\ea \right)
\label{2.10}
\ee
where $\hat{I}$ denotes the $N\times N$ identity matrix.
\par
Let ${\cal I}\ \subset \ {\cal A}$ be the subalgebra of ${\cal A}$,
consisting of matrix-valued sequences obeying homogeneous boundary
conditions. Once equipped with the Lie-product given by the point-wise
commutator:
\[
[X,Y]_n \doteq X_nY_n - Y_n X_n
\]
${\cal I}$ becomes a Lie-algebra and moreover an ideal of
${\cal A}$, and can be identified with its dual through the bilinear form:
\be
(X,Y) = \sum_{n=-\infty}^{+\infty} <X_n, Y_n> \doteq Tr
\sum_{n=-\infty}^{+\infty}X_n Y_n
\label{2.11}
\ee
Our configuration space $M$ will be the affine hyperplane to ${\cal I}$
of matrix-valued sequences obeying (\ref{2.10});  accordingly,
its generic point will be again denoted by $Q$.
\par
Its tangent bundle $T(M)$ will then be the set $\{(Q,K_Q): \ \ Q\in M, K_Q \in
{\cal I}\}$ while the points of its cotangent bundle $T^*(M)$ will be
parametrized as
$\{(Q,\gamma_Q): \ \ Q\in M, \gamma_Q \in {\cal I}^*\simeq {\cal I}\}$.
\par
In the following, for the sake of simplicity, elements of $T(M)$ (resp.
$T^*(M)$) will be shortly denoted by $K_Q$ (resp. $\gamma_Q$).
In the present concrete realisation of the abstract setting introduced in
\cite{R}, the Poisson tensors $\theta_1,\ \ \theta_2$ defined by
(\ref{2.2a},\ref{2.2b}), as linear maps form $T(M)$ and $T^*(M)$, are in
fact linear
operators on ${\cal I}$, and the same is true for the Nijenhuis
tensor ${\cal N}$, given by (\ref{2.3}), which is a linear map form $T(M)$
into itself.
\par
The main result of present Section is contained in the following:
\begin{guess}
The Nijenhuis tensor ${\cal N}$ reduces by restriction on the submanifold
$Im \ \theta_1$.
\label{t1}
\end{guess}
As a consequence, the abstract bi-hamiltonian hierarchy (\ref{2.6})
restricts to a bi-hamiltonian hierarchy of evolution equations on a
one-dimensional lattice.
\par
The theorem (\ref{t1}) will be proved in a constructive way, which will
then also yield the concrete form of the restricted Poisson and Nijenhuis
tensors.
\par
Let us start by evaluating the image of $\theta_1$:
\be
Im \theta_1 = \{ K_Q = \theta_1 \gamma_Q; \gamma_Q \in {\cal I}\}
\label{2.12}
\ee
Let us choose for $K_Q$ and $\gamma_Q$ the parametrisation induced by
(\ref{2.8}):
\be
K_Q=\left( \ba{cc} K_p & <K_q| \\ |K_r> & K_{\hat{Q}}
\ea \right) \ \ ;\ \ \gamma_Q=\left( \ba{cc} \gamma_p & <\gamma_r| \\
|\gamma_q> & \Gamma_{\hat{Q}}\ea \right)
\label{2.13}
\ee
One immediately gets, for $K_Q \in Im\theta_1$, i.e. for $K$ of the form:
\be
K_{Q}= Q \gamma_Q A - A \gamma_Q Q \ \ \ \ \ \ \
(\mbox{for some } \gamma_Q)
\label{2.14}
\ee
the necessary condition:
\be
K_{\hat{Q}}= \hat{0}
\label{2.15}
\ee
Hence, $\hat{Q}$ must be a constant $N\times N$ matrix. In view of the
boundary condition (\ref{2.10}), we have:
\be
\hat{Q}= \hat{I}
\label{2.16}
\ee
In turn, formula (\ref{2.16}) implies:
\[
K_p = <K_q|r> +<q|K_r>
\]
which entails:
\be
p = qr+c \ \ \ \ \ \ \  (c=\mbox{const}.)
\label{2.17}
\ee
Summarizing,
we can assert that:
\begin{enumerate}
\item
under conditions (\ref{2.9}), (\ref{2.10}), $Im \theta_1$ is the
submanifold of ${\cal I}$ consisting of vector fields $K_Q$ such that:
\be
K_{\hat{Q}} =\hat{0} \ \ \ ;\ \ \ K_p = <K_q|r>+<q|K_r>
\label{2.18}
\ee
\item
the set $S\ \subset M$, such that $T(S)=Im\theta_1$ , given by:
\be
S =\{ Q: p=<q|r>, \hat{Q}=\hat{I} \}
\label{2.19}
\ee
is a characteristic leaf of $\theta_1$.
\end{enumerate}
\[ \]
{\footnotesize
We are now ready to show that ${\cal N}$ reduces by restriction on $S$,
namely that $T(S)= Im\theta_1$ is an invariant submanifold for ${\cal N}$.
To this aim, let us introduce the auxiliary vector field $\xi_Q$ through the
formula:
\be
K_Q= (E \xi_Q)A - A\xi_Q
\label{2.20}
\ee
whence it follows, on $T(M)$:
\be
K'_Q \doteq{\cal N}K_Q = (E\xi_q)Q-Q\xi_Q
\label{2.21}
\ee
Parametrizing
$\xi_Q$ as
\be
\xi_Q = \left( \ba{cc} \xi_0 & <\xi_q| \\ |\xi_r> & {\hat{\xi}}
\ea \right)
\label{2.22}
\ee
we get from (\ref{2.20})
\bq
\xi_0&=& (E-1)^{-1}K_p \label{2.23a} \\
<\xi_q|&=& -<K_q| \label{2.23b} \\
|\xi_r>&=& E^{-1}|K_r> \label{2.23c}
\eq
while the matrix $\hat{\xi}$ stays
undetermined.
\par
On the other hand, eq.(\ref{2.21}) yields, for $ Q\in S$:
\bq
K'_p &=& <q|r>(E-1)K_p + <E\xi_q|r> - <q|\xi_r>\label{2.24a} \\
<K'_q|&=& <q|(E\xi_0)+(E-<q|r>) <\xi_q|-<q|\hat{\xi} \label{2.24b} \\
|K'_r> &=& (<q|r>E-1)|\xi_r> -\xi_0|r> +(E\hat{\xi})|r> \label{2.24c}
\eq
By imposing the reduction condition: $\hat{K}'_Q=0$, one gets:
\[
\hat{\xi} = (E-1)^{-1}(-|E\xi_r><q| + |r><\xi_q|)
\]
or, using (\ref{2.23b},\ref{2.23c}):
\be
\hat{\xi} = -(E-1)^{-1}(+|K_r><q| + |r><K_q|)
\label{2.25}
\ee
Then, to establish that $S$ is an invariant submanifold for ${\cal N}$, we
have just to show that:
\be
K_p= <K_q|r> +<q|K_r>
\label{2.26a}
\ee
entails:
\be
K'_p= <K'_q|r> +<q|K'_r>
\label{2.26b}
\ee
and this can be seen  by a direct straightforward (although tedious)
calculation.} $\Box$
\par
Hence, the theorem (\ref{t1}) is proved and moreover we can give the
esplicit form of ${\cal N}|_S$, more precisely we have:
\be
\left[ \ba{c} {\sct |K'_q>} \\ {\sct |K'_r> }\ea  \right]
=\left[ \ba{c} {\sct (<q|r>-E)|K_q> + \left({E \over E-1}(<K_q|r>+<q|K_r>) +
{1 \over E-1}(|r><K_q|+|K_r><q|)\right)|q> }\\
{\sct(<q|r>-E^{-1})|K_r> - \left({1 \over E-1}(<K_q|r>+<q|K_r>) +
{E \over E-1}(|r><K_q|+|K_r><q|)\right)|r> }\ea \right]
\label{2.27}
\ee
The explicit form of $\theta_1|_S$ can be easily obtained by noting that
the points of $S$
(resp. $T(S)$) are completely determined once the $2N$ vector
$(<q|,<r|)^t$
( resp. \linebreak $(<K_q|,<K_r|)^t$) are given.
\par
Hence
points of $T^*(S)$ will be fully characterized by the value of the $2N$
vector ($<\beta_q|,<\beta_r|)$ defined by the duality condition:
\be
Tr\sum_{n=-\ii}^{+\ii} \gamma_{Q,n}K_{Q,n}= \sum_{n=-\ii}^{+\ii}
<\beta _{q_n}|K_{q_n}> +<\beta _{r_n}|K_{r_n}>
\label{2.28}
\ee
which entails:
\[
<\beta_q|= < \gamma_r|- \gamma_p<r|\ \ \ \;\ \ \ \
<\beta_r|= < \gamma_q|- \gamma_p<q|
\]
By direct calculation, one can check that the relation:
\be
 K_Q = \theta _1 \gamma_Q
\label{2.29}
\ee
once restricted to $S$, implies:
\be
\left[ \ba{c} |K_q> \\ |K_r> \ea \right] =
\left(\ba{cc} 0&-\hat{I} \\ \hat{I}& 0\ea\right)
\left[ \ba{c} |\beta_q> \\ |\beta_r> \ea \right] \doteq
J\left[ \ba{c} |\beta_q> \\ |\beta_r> \ea \right] =
   \label{2.30}
 \ee
whence it follows that $\theta_1|_S$ is the canonical Poisson tensor or, in
other words, that the fields variables $|q>, \ |r>$ are endowed with
canonical Poisson brackets:
\be
\ba{c} \{q_j(n), q_k(m)\}_1=\{r_j(n), r_k(m)\}_1=0
 \\ \{q_j(n), r_k(m)\}_1= \delta_{j,k} \delta_{n,m}\ea.
   \label{2.31}
 \ee
It is worthwile
to notice that the restricted
Poisson tensor $\theta_2|_S \doteq {\cal N}|_S \theta_1|_S$  is on the
other hand nonlocal and has a rather cumbersome
form: its non-locality is intimately related with the non degeneracy of
$\theta_1|_S$.
\par
According with the general theory of bi-hamiltonian system, we have still
to show that the restricted vector fields:
\[
\tilde{K}_B = [Q|_S, B]\ \ \ \ \ (\ [B,A]=0\ )
  \]
are invariant vector fields for ${\cal N}|_S$: but this follows by the
invariant
nature of the Lie derivative, provided that $K_B$ belongs to $T(S)$.
\par
Actually, the commutativity condition $[B,A]=0$ implies for matrix $B$ the
form:
\[
B=\left(\ba{cc} b&<0| \\ |0>& \hat{B}\ea\right)
\]
so that, for any $Q\in S$, we have:
\be
\tilde{K}_B=\left(\ba{cc}0 & <q|(b-\hat{B})\\ (\hat{B}-b)|r>
& 0\ea\right)
\label{2.32}
 \ee
So, the family of starting commuting symmetries can be written in terms of
the \linebreak $2N$-vectors:
\be
\left[\ba{c} |K_q^c> \\ |K_r^c> \ea\right]=
\left[\ba{c} \hat{C}^t|q> \\ -\hat{C}|r> \ea\right]
   \label{2.33}
 \ee
$\hat{C}=b-\hat{B}$ being an arbitrary $N\times N$ matrix.
\par
However, we should notice that the resulting evolution
equations:
\be
\left(\ba{c} |q> \\ |r> \ea\right)_{t_l}=[{\cal N}|_S]^{l}
\left(\ba{c} |K_q^c> \\ |K_r^c> \ea\right)
   \label{2.34}
 \ee
will be in general non-local.
Local equations will be get whenever $\hat{C}= c \hat{I}$.
\par
Let us now give some concrete examples of local evolution equations
associated with the linear problem (\ref{2.1}), corresponding to the choice
$\hat{C}=\hat{I}$:
\begin{enumerate}
\item
($l=0$)
\be
\left(\ba{c} |q> \\ |r> \ea\right)_{t_0}=
\left(\ba{c} |q> \\ -|r> \ea\right)=\theta_1|_S
\left(\ba{c} |r> \\ |q> \ea\right)
   \label{2.35a}
 \ee
or, in components:
\be
{\partial q_j \over \partial t_0} = q_j\ \ \ ;\ \ \
{\partial r_j \over \partial t_0}= -r_j   \ \ \ (j=1,....N)
\label{2.35b}
 \ee
with hamiltonian density:
\be
h_0 = -<q|r>= -\sum_{j=1}^N q_j r_j
   \label{2.35c}
 \ee
%****************
\item
($l=1$)
\be
\left(\ba{c} |q> \\ |r> \ea\right)_{t_1}=
\left(\ba{c}( <q|r>- E)|q>  \\  (E^{-1}-<q|r>)|r>\ea\right)
   \label{2.36a}
 \ee
with hamiltonian density:
\be
h_1= -{1 \over 2}<q|r>^2 + <Eq|r>
   \label{2.36b}
 \ee
%*********
\item
($l=2$)
\be
\left(\ba{c} {\sct |q>} \\ {\sctsct |r> }\ea\right)_{t_2}
=-\left(\ba{c} {\sct -E^{2}|q> + E (<q|r>|q>) + (<Eq|r>+<q|E^{-1}r>-
<q|r>^2)|q>}
\\
{\sct \Big(E^{-2}-<Eq|r>+(<q|r>)^2-<q|E^{-1}r>\Big)|r> -
\Big(<q|r>+<E^{-1}q|E^{-1}r>\Big)|E^{-1}r>}
\ea\right)
\label{2.37a}
 \ee
with hamiltonian density:
\be
h_2= -{1 \over 3}<q|r>^3 + <q|r>(<q|E^{-1}r>+<Eq|r>) - <Eq|E^{-1}r>
   \label{2.37b}
 \ee
%******************
\item
($l=-1$)
\be
\left(\ba{c} |q> \\ |r> \ea\right)_{t_1}=[{\cal N}|_S]^{-1}
\left(\ba{c} |q> \\ -|r> \ea\right)=
\left(\ba{c}  -E^{-1}|q>/(1+<E^{-1}q|r>)\\ E|r>/(1+<q|E_r>) \ea\right)
   \label{2.38a}
 \ee
with hamiltonian density:
\be
h_{-1} =\ln (1+<q|E r>)
   \label{2.38b}
 \ee
\end{enumerate}
\par
\[ \]
We will now show that the above geometric reduction procedure is perfectly
equivalent to the peharps more familiar technique, based on the so-called
discrete zero-curvature condition:
\be
U_t + UV - (EV)U=0
\label{2.41}
\ee
obtained by
enforcing compatibility of linear problem (\ref{2.1}) with the auxiliary
linear problem:
\be
{\partial \psi\over \partial t}= V(\lm)\psi
\label{2.40}
\ee
In our case:
\be
U=Q +\lm A
   \label{2.42}
 \ee
where $A$ is given by (\ref{2.9}) and $Q$ belongs to manifold $S$
(\ref{2.19}), i.e.:
\[
Q=\left(\ba{cc} <q|r> & <q| \\ |r> & \hat{I} \ea\right)
\]
To extract from (\ref{2.4}) the hierarchy (\ref{2.34}) we start by
considering the ``stationary equation'' associated with
(\ref{2.41}), namely:
\be
UW-(EW)U=0
\label{2.43}
\ee
Parametrizing $W$ as:
\be
W=\left(\ba{cc} w & <u| \\ |v> & \hat{W}\ea\right)
\label{2.44}
\ee
one gets
\be
\hat{W}   = (E-1)^{-1}\big[|r><u|-|Ev><q|\big]+\tilde{W}(\lm)
\label{2.45a}
\ee
\be
w=-tr\ \hat{W}=   (E-1)^{-1}\big[|r><u|-|Ev><q|\big]-tr\ \tilde{W}(\lm)
\label{2.45b}
\ee
$\tilde{W}(\lm)$ being an arbitrary
\underline{constant} (i.e., \underline{field  independent})
matrix, and the following ``eigenvalue equations'' for $<u|$, $|v>$:
\be
\lm <u| = (E-<q|r>)<u| +<q|(Ew-\hat{W})
\label{2.46a}
\ee
\be
\lm E|v> = (E^{-1}-<q|r>)|Ev> + (w-E \hat{W})|r>
\label{2.46b}
\ee
with $w,\ \hat{W}$ given by (\ref{2.45a}),(\ref{2.45b}).
\par
Eqs. (\ref{2.46a}, \ref{2.46b}) can be solved recursively assuming
$W$, and thus $<u|,\ |v>$, to be given by the following Laurent series in
$\lm$:
\be
\sum_{k=0}^{\ii}W^{(k)}\lm^{-k}
\label{2.47}
\ee
and requiring $\tilde{W}(\lm)$ in (\ref{2.45a}),(\ref{2.45b}),
to be in fact $\lm$-independent.
\par
One gets:
\[
<u^{(0)}|=|v^{(0)}>=0
\]
\be
<u^{(k+1)}|=   (E-<q|r>)<u^{(k)}| +<q|(Ew^{(k)}-\hat{W}^{(k)})
\label{2.48a}
\ee
\be
E|v^{(k+1)}> = (E^{-1}-<q|r>)|Ev^{(k)}> + (w^{(k)}-E \hat{W}^{(k)})|r>
\label{2.48b}
\ee
On the other hand, if $W$ is a solutions of (\ref{2.43}), the same is true
of course for $\lm^{k}W$, where $k$ is any positive integer.\par
Given a two-sided Laurent series on the unit circle:
\[
f(\lm)= \sum_{s=-\ii}^{+\ii} f_{s}\lm^s\ ,
\]
we shall denote, as usual, by $f_+(\lambda)$ (resp. $f_-(\lambda)$) the part
containing only non negative (resp. strictly negative) powers of $\lambda$.
Therefore,  eq. (\ref{2.43}) can be rewritten as:
\be
U(\lm^kW)_+ - (\lm^k E W)_+ U= - U(\lm^k W )_- + (\lm^k E W)_- U
\label{2.50}
\ee
But now, as $U$ is linear in $\lm$, the l.h.s. of (\ref{2.50}) cannot
contain any negative power of $\lm$, while its r.h.s. cannot contain any
strictly positive one.
Hence both sides of (\ref{2.50}) are $\lm$-independent (order  ``zero'' in
$\lm$), and we have:
\be
l.h.s. = Q W^{(k)} - (E W^{(k)})Q
\label{2.51a}
\ee
\be
r.h.s. = -A W^{(k+1)} - (E W^{(k+1)})A
\label{2.51b}
\ee
As far as its $\lm$-dependence is concerned, (\ref{2.51a}) (and of course
\ref{2.51b}) is then \underline{compatible} with $U_t$.
On the other hand it may be readily checked that it belongs indeed to the
manifold $T(S)$, defined in eqs. (\ref{2.18}-\ref{2.19}).
Hence, we can assert that the hierarchy of evolution equations associated
to (\ref{2.1}), (\ref{2.40}) is given by:
\be
{\partial U\over \partial t_k} = - QW^{(k)} + (E W^{(k)}) Q= A W^{(k+1)}-
(EW^{(k+1)})A
\label{2.52}
\ee
They correspond to the following choice for  the matrix $V$ appearing in
formulas (\ref{2.40}), (\ref{2.41}):
\be
V=V_{(k)}=(\lm^kW)_+
\label{2.53}
\ee
and clearly coincide with (\ref{2.34}),
by choosing $-\hat{C}=(tr \tilde{W})\hat{I} + \tilde{W} $(see rqs.
(\ref{2.33}),
(\ref{2.45a}),(\ref{2.45b})). In particular, $\hat{C}=\hat{I}
\leftrightarrow \tilde{W}= - { 1\over N + 1} \hat{I}$. Indeed,
looking at (\ref{2.52}), (\ref{2.22}),
we see that $W$ is the generating function of the fields $\xi_Q$.

\section{Direct and inverse problem}
\label{sec3}

In this Section, we outline the solution of the direct and inverse problem
associated to (\ref{2.1}); for simplicity, we restrict considerations to
the $2\times 2$ matrix case, when we have:
\be
U_n(\lm)=Q_n +\lambda A = \left(\ba{cc} \lambda+q_nr_n & q_n \\ r_n & 1 \ea
\right)
\label{3.1}
\ee
As explained in section $2$, $\{q_n\}_{n\in Z}$, $\{r_n\}_{n\in Z}$
are real valued sequences vanishing at $\pm \infty$
\be
\lim_{|n|\rightarrow \infty}q_n = \lim_{|n|\rightarrow \infty}r_n =  0
\label{3.2}
\ee
For the linear problem:
\be
\psi_{n+1}= U_n\psi_n
\label{3.3}
\ee
we can naturally define the transfer matrix:
\bq
T_{n,m}(\lm)=U_{n-1}........U_{m}\ \ \ \ (n\geq m)
\label{3.4a} \\
T_{n,m}(\lm)=U_{n}^{-1}U_{n+1}^{-1}........U_{m-1}^{-1}\ \ \ \ (n< m)
\label{3.4b}
\eq
such that:
\be
\psi_n = T_{n,m} \psi_m
\label{3.5}
\ee
We can then introduce the ``{\it Jost matrices}'':
\bq
T_{n}^{-}(\lm)\doteq \lim_{m\fd-\ii}T_{n,m}(\lm)E_{m}(\lm)=
(\phi_n(\lm),\tilde{\phi}_n(\lm))
\label{3.6} \\
T_{n}^{+}(\lm)\doteq \lim_{m\fd+\ii}T_{n,m}(\lm)E_{m}(\lm)=
(\tilde{\phi}_n(\lm),\psi_n(\lm))
\label{3.7}
\eq
where:
\be
E_n(\lm)\doteq \left(\ba{cc} \lm^n & 0 \\ 0 & 1 \ea \right)
\label{3.8}
\ee
is a fundamental matrix solution of the asymptotic (or ``undressed'')
problem:
\be
E_{n+1}(\lm)= \left(\ba{cc} \lm & 0 \\ 0 & 1 \ea \right) E_{n}(\lm)
\label{3.9}
\ee
and $\varphi_n,\ \  \tilde{\varphi}_n, \ \ \psi_n , \ \ \tilde{\psi}_n$
are $2$-column vector solutions of (\ref{2.1}), the ``{\it Jost solutions}''.
\par
Cleary, the asymptotic solution $E_n(\lm)$ (\ref{3.8}) is bounded on
the unit circle $|\lm|=1$, which will then be the continuous spectrum of
(\ref{2.1}), (\ref{3.2}).
\par
On the unit circle, the monodromy matrix is then defined as:
\be
T(\lm)= T^+_n(\lm) [T_n(\lm)]^{-1}
\label{3.10}
\ee
In the following, we shall call ``{\it spectral parameters}'' the elements
of monodromy matrix:
\be
T(\lm)= \left(\ba{cc} a(\lm) & \tilde{b}(\lm) \\ b(\lm) &
\tilde{a}(\lm) \ea \right)
\label{3.11}
\ee
As $det\ U_n=\lm$, we have:
\be
det\ T_{n,m}(\lm)=\lm^{n-m},\ \ \ \ \ \det\ T_{n}^{\pm}(\lm)=\lm^{n}
\label{3.12a} \\
\ee
so that:
\be
det\ T(\lm)=1
\label{3.12b}
\ee
Formulas (\ref{3.12a}) implies that $\varphi_n$, $\tilde{\varphi}_n$ and
$\psi_n$, $\tilde{\phi}_n$ are two pairs of independent vector solutions
of (\ref{2.1}) on the unit circle $(|\lm|=1)$, while formulas
(\ref{3.11}), (\ref{3.12b}) entail, on the unit circle:
\be
a(\lm) \tilde{a(\lm)}= 1+b(\lm)\tilde{b}(\lm)
\label{3.12c}
\ee
\[ \]
{\it Direct problem}
\par
The direct problem amounts to determine the monodromy matrix (\ref{3.11})
once the fields $\{q_n\},\ \ \{r_n\}$ are given.
\par
To this aim, it is convenient to introduce the normalized
vector sequences:
\be
\tilde{\chi}\doteq \lm^{-n} \tilde{\psi}_n;\ \ \ \ \
{\varphi}\doteq \lm^{-n} \phi_n
\label{3.13a}
\ee
such that:
\be
\lim_{n\fd -\ii} \tilde{\chi}_n(\lm)= \left( \ba{c} 1\\0 \ea \right)=
\lim_{n\fd +\ii} \varphi_n(\lm)
\label{3.13b}
\ee
Somewhat loosely, in the following $\tilde{\chi}_n$, $\varphi_n$ will be
denoted as ``{\it Jost solution}'' as well.
\par
The spectral parameters are naturally defined as Wronskians of independent
vector solutions; namely we have:
\be \ba{l}
a(\lm) = W(\varphi,\psi) \\
\tilde{a}(\lm) = W(\tilde{\phi},\tilde{\chi}) \\
b(\lm) = \lm^n W(\tilde{\chi},\varphi) \\
\tilde{b}(\lm) = \lm^{-n} W(\tilde{\phi},\psi)
\ea \label{3.14}
\ee
where $W(a,b)$ is the determinant of the matrix whose columns are the
$2$-vectors $a$ and $b$.
\par
It easily seen that the vector sequences $\tilde{\chi}_n,\ \ \psi_n,
\ \ \varphi_n, \ \ \tilde{\varphi}_n$ satisfy the following ``discrete
integral'' equations:
\be
\tilde{\chi}_n(\lambda)  =  \left( \ba{c} 1 \\0 \ea \right) +
        \sum_{k=n}^{\infty} \left( \ba{cc}
        \lambda^{-1} & 0 \\ 0 & \lambda^{k-n} \ea \right)
        \left( \ba{cc} q_kr_k & q_k \\ r_k & 0\ea \right) \tilde{\chi}_k (
\lambda)
\label{3.15a}
\ee
\be
\psi_n(\lambda) = \left( \ba{c} 0 \\ 1 \ea \right) + \sum_{k=n}^{\infty}
\left( \ba{cc}
\lambda^{n-(k+1)} & 0 \\ 0 & 1 \ea \right)
\left( \ba{cc}
q_kr_k & q_k \\ r_k & 0\ea \right) \psi_k (\lambda)
\label{3.15b}
\ee

\be
\varphi_n(\lambda)= \left( \ba{c} 1 \\0 \ea \right)-
\sum_{k=-\infty}^{n-1} \left(
\ba{cc} \lambda^{-1}  & 0 \\ 0 & \lambda^{k-n}
 \ea \right)
\left( \ba{cc}
q_kr_k & q_k \\ r_k & 0\ea \right)
\varphi_k(\lambda)
\label{3.15c}
\ee
\be
\tilde{\phi}_n(\lambda)= \left( \ba{c} 0 \\ 1 \ea \right)-
\sum_{k=-\infty}^{n-1} \left(
\ba{cc}
\lambda^{n-k-1} & 0 \\ 0 & 1 \ea \right)
\left( \ba{cc}
q_kr_k & q_k \\ r_k & 0\ea \right)
\tilde{\phi}_k(\lambda)
\label{3.15d}
\ee
The analyticity properties of the Jost solutions are summarized by the
following:
\begin{guess}

If the sequences $\{q_n\}$, $\{r_n\}$ are such that:
\be
\lim_{|n|\fd\ii} n^2 q_n=\lim_{|n|\fd\ii} n^2 r_n=0
\label{3.16}
\ee
then $\psi_n$, $\varphi_n$ are analytic functions
of $\lm$ in the domain $|\lm|>1$ and are continuously differentiable for
$|\lm|\geq 1$; analogously, $\tilde{\chi}_n,\ \ \tilde{\varphi}_n$
are analytic functions of $\lm$ for $|\lm| <1$ and continuously differentiable
for $|\lm|\leq 1$.
\label{t2}
\end{guess}
We outline the proof of theorem (\ref{t2}) for $\varphi_n$.
\par
 {\footnotesize
First of all, we equip $\bf{C}^N$ with the norm:
\[{
\|x\|= \max_{k=1...N}|x_k|}
\]
 so that linear transformations in $\bf{C}^N$ are naturally
equipped with the norm:
\[{
\|X\|= \max_{j=1...N}\ \sum_{j=1}^{N} |x_{ij}| }
\]
 Then, the following inequality holds for (\ref{3.15c}):
\be{
\left\|
 \left( \ba{cc} \lm^{-1} & 0 \\ 0 & \lm^{k-n} \ea \right) U_{1,k}
\right\|
\leq \|U_{1,k}\|  \ \ \ \ \ (|\lm| \geq 1)}
\label{3.18a}
\ee
 where
\be{
U_{1,k} = \left( \ba{cc} q_k r_k & q_k \\ r_k & 0\ea \right)}
\label{3.18b}
\ee
 So, writing the Neumann series solutions of (\ref{3.15a}):
\be{
\varphi_n(\lm)= \sum_{l=0}^{\ii}(-1)^l F_n^{(l)}
\left( \ba{c} 1 \\ 0 \ea \right)}
\label{3.19}
\ee
 where:
\[{
F_n^{(0)} = \left( \ba{cc} 1 & 0 \\ 0 & 1 \ea \right)}
\]
\be{
F_n^{(l)}(\lm) = \sum_{k_1>n} G(n,k_1)\sum_{k_2>k_2} G(k_1,k_2)......
\sum_{k_2>k_2} G(k_{l-1},k_l)\ \ \ \ \ (l>0)}
\label{3.20a}
\ee
\be{
G(n,k)= \left( \ba{cc} \lm^{-1} & 0 \\ 0 & \lm^{k-n} \ea \right)U_{1,k}}
\label{3.20b}
\ee
The inequality (\ref{3.18a}) implies, by iteration:
\be{
\|F_n^{(l)}(\lm) \|\leq {1 \over l!} \left( \sum_{k>n} \|U_{1,k}\|\right)^l
\ \ \ \ \ \ \ l\geq 0}
\label{3.21}
\ee
 Therefore:
\be{
\|\varphi(\lm)\|\doteq \sup_{ n \in Z}
\|\varphi_n(\lm)\|\leq \exp \gamma}
\label{3.22}
\ee
 where:
\be{
\gamma = \sum_{n \in Z} \|U_{1,n}\|}
\label{3.23}
\ee
 Hence, if $\gamma<\ii$, $\varphi_n(\lm)$ is analytic for $|\lm|>1$.
On the other hand:
\[ {
\|U_{1,n}\|= \max \left\{ |q_n| \big[1+|r_n|\big], |r_n| \right\}
\leq |q_n| +|r_n| + 1/2 \big[|q_n|^2+ |r_n|^2 \big]
}\]
so that the existence of $\gamma $ is guaranteed
whenever the sequences $\{q_n\},\ \ \{r_n\}$ belong to $l_1$.
\par
 A similar procedure leads to the following result for
${\partial \varphi_n \over \partial \lm}$ :
\be{
\left\|{\partial \varphi_n \over \partial \lm}\right\|\leq a+|n|b \ \ \
|\lm|\geq 1}
\label{3.24}
\ee
 which holds, with suitable coefficients $a$ and $b$,
whenever:
\be{
\sum_{n \in Z} (1+|n|) \|U_{1,n}\|<\ii}
\label{3.25}
\ee
Hence, provided $q_n$ and $r_n$ vanish faster than $n^{-2}$ as
$|n|\fd\ii$, $\varphi_n(\lm)$ is continuously differentiable with
respect to $\lm$ for $|\lm|\geq1$. }
$\Box$
\par
We can thus assert that, whenever $\{q_n\}$, $\{r_n\}$ belong to
$l_1$, the diagonal entries of the monodromy matrix
$a(\lm)$ and $\tilde{a}(\lm)$ ({3.14})
are analytic respectively outside and inside the unit circle.
Morever, due to their asymptotic behaviour in $\lm$:
\be
\lim_{|\lm|\fd \ii} a(\lm)=\lim_{|\lm|\fd \ii}  det (\varphi,\psi)=
det \left(\ba{cc} 1&0\\ 0&1\ea\right)=1
\label{3.25bisa}
\ee
\bq
\lim_{|\lm|\fd 0} \tilde{a}(\lm)=\lim_{|\lm|\fd \ii}  det (
\tilde{\varphi},\tilde{\chi})&=&
\prod_{\sr{j\neq n}{j=-\ii}}^{+\ii} (1+r_jq_{j-1})
det \left(\ba{cc} q_{n-1} & 1 \\ 1 & -r_n \ea\right)=\\
&=&- \prod_{j=-\ii}^{+\ii} (1+r_jq_{j-1})
\label{3.25bisb}
\eq
both $a$ and $\tilde{a}$
have at most a finite number of zeros, say $N$ and $\tilde{N}$
respectively, in their analyticity domains.
These zeroes will be denoted as $\{\lm_j\}_{j=1}^N$ and
$\{\tilde{\lm}_j\}_{j=1}^{\tilde N}$ and will be assumed to be simple.
\par
If in addition the stronger condition (\ref{3.25}) is satisfied, then
the entries of the monodromy matrix (\ref{3.11}) are H\"older
continuous on the unit circle (see again (\ref{3.14})).
Morever, the analyticity properties
of $a(\lm)$ and $\tilde{a}(\lm)$ imply that the scalar Riemann problem on
the unit circle (\ref{3.12c}) can be solved through the formulas:
\bq
a(\lambda)={\prod^{N}_{j=1}(\lambda-\lambda_j) \over
           \prod^{\tilde{N}}_{j=1}(\lambda-\tilde\lambda_j)}
                          \lambda^{\sigma}
          \exp\left[ {1 \over 2\pi i} \oint_{|z|=1}{\ln\Big(1+b(z)
          \tilde b(z)\Big)- \sigma \ln z
                         \over z - \lambda} dz \right]
\label{3.25trisa}
\\
\mbox{for }|\lambda| > 1 \nonumber
\eq
\bq
\tilde a(\lambda)={\prod^{\tilde{N}}_{j=1}(\lambda-\tilde\lambda_j)
                 \over \prod^{N}_{j=1}(\lambda-\lambda_j)}
                        \exp\left[{1 \over 2\pi i} \oint_{|z|=1}
                     {\ln\Big(1+b(z)\tilde b(z)
            \Big)- \sigma\ln z \over z-\lambda}
                           dz \right]
\label{3.25trisb}
\\
\mbox{for }|\lambda| < 1 \nonumber
\eq
where $\sigma$ is the index of the Riemann problem, i.e. the variation of
$Arg\big(1+b(\lm)\tilde{b}(\lm)\big)$ after a cycle; condition (\ref{3.25bisa})
clearly implies $\sigma =N-\tilde{N}$.
\par
In the following, we shall \underline{assume} the index $\sigma$ to be
zero so that $N=\tilde N$: this is certainly true in the reflectionless case
($b=\tilde{b}=0$).

\[ \]
{\it Inverse problem}
\par
The inverse problem amounts to reconstruct the sequences $\{q_n\}$,
$\{r_n\}$, once the monodromy matrix is given.
\par
To solve it, we rewrite eq.(\ref{3.10}) in terms of the appropriate column
vector solutions:
\bq
{\tilde{\psi}_n \over \tilde{a}}= \psi_n +
{b \over \tilde{a}}\lm^n \tilde{\chi}_n
\label{3.26a} \\
  \\
{{\varphi}_n \over {a}} = \tilde{\chi}_n + {b \over {a}} \lm^{-n}\psi_n
\label{3.26b}
\eq
Under the above analiticity conditions on $\tilde{\phi}$, $\psi$,
$\varphi$, $\tilde{\chi}$, eqs. (\ref{3.26a}, \ref{3.26b}) are two vector
{\it Riemann-Hilbert} problems, which can be solved (i.e. reduced to
singular integral equations) through the well-known Plemelj formulas.
One gets:
\be
\psi_n(\lambda)= \left( \ba{c} 0 \\ 1 \ea \right) + {1 \over 2 \pi i }
              \oint \tilde{\rho}(\zeta) { \zeta^n\tilde{\chi}_n(\zeta) \over
                    \zeta-\lambda} d\zeta - \sum_{k=1}^{N}
                {\tilde{\lambda}_k^n\tilde{\gamma}_k\tilde{\chi}
                (\tilde{\lambda}_k)
                 \over (\tilde{\lambda}_k-\lambda)}
\label{3.27a}
\ee
\be
\tilde{\chi}_n(\lambda)= \left( \ba{c} 1 \\ 0 \ea \right) - {1 \over 2 \pi i }
              \oint \rho(\zeta) { \zeta^{-n}\psi_n(\zeta) \over
                    \zeta-\lambda} d\zeta - \sum_{k=1}^{N}
                {\lambda_k^{-n}\gamma_k\psi(\lambda_k) \over
(\lambda_k-\lambda)}
\label{3.27b}
\ee
where we have set:
\be
\rho(\lm)=b(\lm)/a(\lm);\ \ \ \ \ \tilde{\rho}(\lm)=\tilde{b}(\lm)/
\tilde{a}(\lm)
\label{3.28a}
\ee
\be
\gamma_k=b_k/a'(\lm_k);\ \ \ \ \ \tilde{\gamma}_k=\tilde{b}_k/\tilde{a}'(
\tilde{\lm}_k)
\label{3.28b}
\ee
In formulas (\ref{3.28b})
 $b_k$ ( resp.$\tilde{b}_k$) are the ratios between $\phi$ and $\psi$ (resp.
$\tilde{\phi}$ and $\tilde{\psi}$) at $\lm=\lm_k$ (resp. $\tilde{\lm}=
\tilde{\lm}_k$).\par
Of course, in the reflectionless case
formulas (\ref{3.27a},\ref{3.27b}) yield a system of $2N$
linear algebraic equations for $\psi_n(\lm_k)$,
$\tilde{\chi}_n(\tilde{\lm}_k).$ \par
Finally, as usual, the sequences $\{q_n\}, \ \{r_n\}$ are given in terms of
the leading terms of the asymptotic behaviour
of the vector solutions $\psi_n$, $\tilde{\chi}_n$.
For instance, taking into account that:
\[
\psi_n = \left( \ba{c} 0 \\ 1 \ea \right) + \lm^{-1} \left( \ba{c}
\psi_{1,n}^{(1)} \\ \psi_{2,n}^{(1)} \ea \right) + O(\lm^{-2})
\]
we have:
\bqw
q_n &=& - \psi_{1,n}^{(1)} \\
  \\
r_n &=& {(\psi_{2,n+1}^{(1)} -\psi_{2,n}^{(1)} )\over \psi_{1,n}^{(1)} }
\eqw
\par
To find the time-evolution of the spectral data corresponding to eq.
(\ref{2.41}), (\ref{2.40}), we notice that, if the matrix
$V$ appearing in the ausiliary linear problem (\ref{2.40}) is given by
(\ref{2.53}), the monodromy matrix undergoes
the time evolution:
\be
{ \partial T(\lm) \over \partial t_k} = [ \bar{V}_{(k)}(\lm), T(\lm)]
\label{3.29}
\ee
where:
\be
\bar{V}_{(k)}(\lm)= \lim_{|n|\fd\ii} {V}_{(k)\ n}(\lm)= \lm^k
\left(\ba{cc} - tr\ \tilde{W} & 0\\ 0 & \tilde{W} \ea\right)
\label{3.30}
\ee
Restricting considerations to the $2\times 2$ case, we can thus write:
\be
\bar{V}_{(k)}(\lm) = c \lm^k \sigma_3
\label{3.31}
\ee
where $c$ is an arbitrary scalar constant and $\sigma_3$ is the usual
Pauli matrix.
\par
Consequently we have:
\bq
{\partial a\over \partial t_k}=&{\displaystyle
{\partial \tilde a \over \partial t_k}}&=0
\label{3.32a}
\\
{\partial b\over \partial t_k}=-2c  \lm^k b\ \ \ &;& \ \ \
{\partial \tilde b \over \partial t_k}= 2c  \lm^k \tilde b
\label{3.32b}
\eq
of course, eqs. (\ref{3.32a}) imply that $\lm_r,\ \ \tilde{\lm}_r$ are
constant in time for any equation of the hierarchy (``isospectral
deformation''), while the normalization coefficient $\gamma_r$,
$\tilde{\gamma}_r$ evolve according to equations:
\be
{\partial{\gamma}_r\over \partial t_k}=-2(\lm_r)^k \gamma_r\ \ \ ; \ \ \
{\partial \tilde{\gamma} \over \partial t_k}= 2(\tilde{\lm}_r)^k \tilde\gamma_r
\label{3.33}
\ee

\section{r-matrix and action-angle variables}
\label{sec4}

We have seen in sec. \ref{sec2} (formulas (\ref{2.30}),(\ref{2.31}))
that our hierarchy of discrete evolution equations are hamiltonian with
respect to the canonical Poisson  tensor or, in other words, that the fields
variables $|q>$, $|r>$ are endowed with the canonical Poisson bracket.
\par
As consequence, we have the ``ultra-local'' Poisson bracket relation
\cite{librogiallo}:
\be
\{ U_n(\lm) \sr{\otimes}{,} U_m(\mu)\}=
[r(\lm,\mu), U_n(\lm)\otimes U_m(\mu)  ]\delta_{n,m}
\label{4.1}
\ee
Restricting again considerations to $2\times 2$ matrices $U_n$, we have for
$4 \times 4$ r-matrix $r(\lm,\mu)$ the formula:
\be
r(\lm,\mu)=r(\lm-\mu)={1 \over \lm-\mu}\left(\ba{cccc} 1 & 0 & 0 & 0\\
0& 0 & 1 & 0   \\  0& 1 & 0 & 0   \\   0 & 0 & 0 & 1\ea\right)
\label{4.2}
\ee
i.e. the same expression as for the NLS case.
\par
{}From (\ref{4.1}) one easily gets the analogous relation for the transfer
matrix, and finally the Poisson-bracket relation for the monodromy matrix,
that reads:
\be
\{T(\lm) \sr{\otimes}{,} T(\mu)\}= r^+  (\lm,\mu)[T(\lm)\otimes T(\mu)]-
[T(\lm) \otimes T(\mu)] r^-(\lm, \mu)
\label{4.3}
\ee
where:
\[
r^{\pm}(\lm,\mu)= \lim_{n\fd\ii}(E^{-1}(n,\lm)\otimes E^{-1}(n,\mu)
r(\lm,\mu)E(n,\lm)\otimes E(n,\mu)
\]
that is:
\be
r^{\pm}=              \left( \ba{cccc} {1 \over \lambda-\mu} & 0 & 0 & 0 \\
                               0 & 0 & \mp i \pi \delta(\lambda-\mu) & 0 \\
                       0 &  \pm i \pi \delta(\lambda-\mu) & 0 & 0 \\
                       0 & 0 & 0 & {1 \over \lambda-\mu} \ea \right)
\label{4.4}
\ee
In formula (\ref{4.4}), ${1 \over \lm-\mu }$ denotes the principal-value
distribution, and we have used the distribution formula:
\be
\lim_{n\fd\pm\ii} {(\mu/\lm)^n\over \lm-\mu}= \mp i \pi \delta (\lm-\mu)
\label{4.5}
\ee
Eq. (\ref{4.3}) implies the following Poisson-brackets for spectral
parameters:
\bq
\{ a(\lambda), a(\mu)\} &=&\{ a(\lambda), \tilde{a}(\mu)\}
=\{ \tilde{a}(\lambda), \tilde{a}(\mu)\} = 0
\label{4.6a}\\
\{ b(\lambda), b(\mu)\}
&=& \{ \tilde{b}(\lambda), \tilde{b}(\mu)\} = 0
\label{4.6b}\\
\{ b(\lambda), \tilde{b}(\mu)\} &=& 2 \pi i \delta(\lambda-\mu)\
\tilde{a}(\mu) a(\mu)
\label{4.6c}\\
\{  a(\lambda), b(\mu) \} &=& - \{ b(\mu),a(\lambda)\} =- {1 \over \lm-\mu}
a(\lambda) b(\mu) -i\pi \delta(\lambda-\mu)\  b(\mu)a(\mu)
\label{4.6d}
\eq

We can thus construct canonical spectral variables, given by:
\be
\ba{lr} \alpha(\lm)\doteq \ln a(\lm)\tilde{a}(\lm)  \ \ \ & \ \ \
\beta(\lm)\doteq {1\over 4\pi i}\ln {\tilde{b}(\lm)\over {b}(\lm)}
\\
\lm_j &     \nu_j\doteq \ln \gamma_j
\\
\tilde\lm_j &     \tilde\nu_j\doteq \ln \tilde\gamma_j
\ea
\label{4.7}
\ee
obeying the Poisson bracket realtions:
\be
\ba{c}
\{\alpha(\lm),\beta(\mu)\}=\delta (\lm-\mu)
 \\
\{\lm_j,\nu_k\} =\{\lm_j,\nu_k\} =\delta_{jk}\ea
\label{4.8}
\ee
(all other P.B. being identically zero).
\par
As usual, in terms of the spectral variables, the continuous and the
discrete spectrum contributions are separated. Moreover,
$\alpha(\lm),\ \{\lm_j\},\ \{\tilde\lm_j\}$ are constant along any flow of
the hierarchy, while $\beta(\lm),\ \{\nu_j\},\ \{\tilde\nu_j\}$
evolve linearly.
\par
The ``action'' variable $\alpha(\lm)$ is defined only on the unit circle.
However, it can be expressed as a sum of two functions having a
single-valued analytic branch for $|\lm|>\max_j|\lm_j|$ and
$|\lm|<\min_j|\tilde{\lm_j}|$ respectively, uniquely
defined by the asymptotics (\ref{3.25bisa}),(\ref{3.25bisb}).
Such branches will be denoted as $\ln a(\lm)$ and $\ln \tilde{a}(\lm)$
respectively.
Then formulas (\ref{3.25trisa}),(\ref{3.25trisb})
imply for such branches the following power series expansions:
\be
\ln a(\lm)=\sum_{k=1}^{\ii}\lm^{-k}J_k
\label{4.9a}
\ee
\be
\ln\tilde{a}(\lm)=\sum^{\ii}_{k=0}\lm^{k}\tilde{J}_k
\label{4.9b}
\ee
where:
\be
J_k=-{1 \over 2\pi i}\oint_{|z|=1}z^{k-1}\alpha(z)dz+\sum_{j=1}^N{1 \over k}
(\tilde{\lm}_j^k-\lm_j^k)
\label{4.10a}
\ee
\be
\left\{\ba{lr}
{\displaystyle
\tilde{J}_k={1 \over 2\pi i}\oint_{|z|=1}{1 \over z^{k+1}}\alpha(z)dz+
\sum_{j=1}^N{1 \over k}(\lm^{-k}_j-\tilde{\lm}^{-k}_j)} & (k\neq 0) \\
{\displaystyle
\tilde{J}_0={1 \over 2\pi i}\oint_{|z|=1}{1 \over z}\alpha(z)dz+
\sum_{j=1}^N (\ln \tilde{\lm}_j-\ln \lm_j) }& (k = 0)
\ea \right.
\label{4.10ba}
\ee
In the following, we will show that the evolution equations (\ref{2.52})
are generated by the hamiltonians $J_k$, i.e.:
\be
{\partial U_n(\lm) \over \partial t_k}=\{J_k,U_n(\lm)\}
\label{4.11}
\ee
Indeed, from the Poisson-bracket relation (\ref{4.1}), we obtain:
\bq
\{ trT(\mu),U_n(\lm)\}
&=&{1 \over \lm-\mu}\big(V_{n+1}(\mu)(\lm A +Q)-(\lm A+Q)V_n(\mu) \big)=\\
\label{4.12aaaa}
&=&V_{n+1}(\mu)A- AV_n(\mu)+
{1 \over \lm-\mu}\big({V}_{n+1}(\mu)U(n,\mu)-U(n,\mu){V}_n(\mu)\big)
\nonumber
\eq
where $V_n$ is defined in terms of the Jost solutions:
\be
V_n(\mu)=\left(\ba{cc} \psi_{2,n}(\mu)\varphi_{1,n}(\mu)&
-\psi_{1,n}(\mu) \varphi_{1,n}(\mu)
\\ \psi_{2,n}(\mu)\varphi_{2,n}(\mu)& -\psi_{1,n}(\mu)\varphi_{2,n}(\mu)
\ea\right)+
\left(\ba{cc} -\tilde{\chi}_{2,n}(\mu)\tilde{\phi}_{1,n}(\mu)
&\tilde{\chi}_{1,n}(\mu)\tilde{\phi}_{1,n}(\mu)
\\ -\tilde{\chi}_{2,n}(\mu)\tilde{\phi}_{2,n}(\mu)
&\tilde{\chi}_{1,n}(\mu)\tilde{\phi}_{2,n}(\mu) \ea\right)
\label{4.12b}
\ee
It can be immediately seen that
\[
\big({V}_{n+1}(\mu)U(n,\mu)-U(n,\mu){V}_n(\mu)\big)=0
\]
Then formula ($134$) entails:
\[
\{a(\mu),U_n(\lm)\}=V^{(+)}_{n+1} A-AV_n^{(+)}
\]
\[
\{\tilde{a}(\mu),U_n(\lm)\}=V^{(-)}_{n+1} A-AV_n^{(-)}
\]
where $V^{\pm}$ are  of course the projections of $V$ outside and inside
the unit circle. Hence:
\be
\{\ln a(\mu),U_n\}=\bar{W}^{(+)}_{n+1} A-A\bar{W}_n^{(+)}
\label{4.13a}
\ee
\be
\{\ln \tilde{a}(\mu),U_n\}=\bar{W}^{(-)}_{n+1} A-A\bar{W}_n^{(-)}
\label{4.13b}
\ee
where
\be
\bar{W}^{(\pm)}= \left\{\ba{c} a^{-1}V^{(+)} \\\tilde{a}^{-1}V^{(-)}
\ea\right.
\label{4.14}
\ee
are again analytic functions of $\lm$ for $|\lm|>\max_j|\lm_j|$ and
$|\lm|<\min_j |\tilde{\lm}_j|$ respectively; they obey the
asymptotic conditions:
\be
\lim_{|\lm|\fd\ii} \bar{W}^{(+)}=\left(\ba{cc} 1 & 0\\ 0 & 0\ea\right)
\label{4.15a}
\ee
\be
\lim_{|\lm|\fd 0} \bar{W}^{(-)}={1 \over 1+r_nq_{n-1}}
\left(\ba{cc} r_n q_{n-1}& q_{n-1}\\ r_n& 1\ea\right)
\label{4.15b}
\ee
\be
\lim_{n\fd\ii} \bar{W}^{(+)}=\left(\ba{cc} 1 & 0\\ 0 & 0\ea\right)
\label{4.15c}
\ee
\be
\lim_{n\fd -\ii} \bar{W}^{(-)}=\left(\ba{cc} 0 & 0\\ 0 & -1\ea\right)
\label{4.15d}
\ee
Formulas (\ref{4.13a}),(\ref{4.13b}) yield:
\be
\{J_k,U_n\}= \bar{W}^{(+)(k)}_{n+1} A-A\bar{W}_n^{(+)(k)}
\label{4.16a}
\ee
\be
\{\tilde{J}_k,U_n\}= \bar{W}^{(-)(k)}_{n+1} A-A\bar{W}_n^{(-)(k)}
\label{4.16b}
\ee
On the other hand, $\bar{W}^{(+)}$, defined by (\ref{4.14}) and $W$,
defined by (\ref{2.43})(\ref{2.47}),
  obey the same stationary equation (\ref{2.43})
and have the same analyticity properties with respect to $\lambda$.
Moreover, their asymptotic behaviours differ essentially just for a
constant multiple of the identity matrix, which plays no role in the
recursion relation: hence  formula (\ref{4.16a}) coincides with
(\ref{2.52}).
\par
We end this section by noting that formula (\ref{4.1})
does hold even in the $(N+1)\times (N+1)$ matrix case. For this general
spectral problem, the $r$-matrix is given by:
\be
r = {1 \over \lambda-\mu}\left(\ba{cccc}R^{1}_{1}  & ...&...& R^{N+1}_{1}
\\ ...&...&...&...\\...&...&...&...\\
R^{1}_{N+1}  & ...&...& R^{N+1}_{N+1} \ea\right)
\label{4.17}
\ee
where
\be
(R^{l}_{m})_{ij}=\delta_{il}\delta_{mj}\ \ \ \ \ \ (i,j=1,....N+1)
\ee

\section{Continuum limit}
\label{sec5}

\begin{guess}
In the continuum limit: $h\fd 0,\; n\fd 0,$
$x=nh$ finite,\\
 with the rescalings:
\be
\ba{lcr}  q\fd h\,q &,& r\fd h\,r\ea
\label{5.1}
\ee
we have :
\be
\lim_{h\fd 0}{{\cal N} +I \over h}=\Lambda
\label{5.2}
\ee
where $\Lambda$ is the recursion operator of the vector AKNS hierarchy.
\label{t3}
\end{guess}

%%%%%%%%%%%%%%%%%%
In fact, from eq.(\ref{2.27}):
\be \left\{ \ba{l}
{\sct {K'}_q =  (<q|r>-E)|K_q> + \left({E \over E-1}(<K_q|r>+<q|K_r>) +
{1 \over E-1}(|r><K_q|+|K_r><q|)\right)|q> }\\
{\sct{K'}_r=(<q|r>-E^{-1})|K_r> - \left({1 \over E-1}(<K_q|r>+<q|K_r>) +
{E \over E-1}(|r><K_q|+|K_r><q|)\right)|r> }\ea \right.
\label{5.3}
\ee
by noting that:
\bq
E= e^{h\partial_x} = 1+h\partial_x+O(h^2)\nonumber\\
\label{5.4}\\
E^{-1}= e^{-h\partial_x} = 1-h\partial_x+O(h^2)\nonumber\\
\eq
and thus
\be
\ba{lcr} {E \over E-1}= {1 \over h}\partial^{-1}_x +O(1)& ; &
{1 \over E-1}= {1 \over h}\partial^{-1}_x +O(1)\ea
\label{5.6}
\ee
we obtain:
\be \left\{ \ba{l}
{\sct {K'}_q= -K_q + h[-\partial_x K_q +  {\partial_x}^{-1}(<K_q|r>+<q|K_r>+
|r><K_q|+|K_q><q|)]|q>+O(h^2)}\\
\\
{\sct {K'}_r= -K_r + h[\partial_x K_r - {\partial_x}^{-1}(<K_q|r>+<q|K_r>+
|r><K_q|+|K_q><q|)]|r>+O(h^2)}
\ea \right.
\label{5.7}
\ee
so that:
\be
\lim_{h\fd 0}{{\cal N} +I\over h}=
\left[\ba{cc}
{\sct -\partial_x + {\partial_x}^{-1}\Big(<\cdot|r>+|r><\cdot|\Big)|q> }&
{\sct - {\partial_x}^{-1}\Big(<q|\cdot>+|\cdot><q|\Big)|q>}\\
{\sct - {\partial_x}^{-1}\Big(<\cdot|r>+|r><\cdot|\Big)|r>} &
{\sct \partial_x - {\partial_x}^{-1}\Big(<q|\cdot>+|\cdot><q|\Big)|r>}
\ea \right]
\label{5.8}
\ee
which is the recursion operator for the vector AKNS case \cite{14}.
In particular, in the $2 \times 2$ matrix case, we get the familiar
recursion operator of the standard AKNS hierarchy:
\be
\lim_{h\fd 0}{{\cal N} +I\over h}\ = \ \left[\ba{cc}-\partial_x
+2q\int^x_{-\ii}r\cdot & 2q\int^x_{-\ii}q\cdot\\
-2r \int^x_{-\ii}r\cdot & \partial_x-2r\int^x_{-\ii}q\cdot\ea\right]
\label{5.9}
\ee
For instance, by taking $\left({{\cal N} +I \over h} \right)^{2}$
we get the equations:
\be \left\{\ba{l}
{\textstyle q_t=q_{xx}-2q^2r} \\
\\
{\textstyle r_t=-r_{xx}+2r^2q }
\ea \right.
\label{5.10}
\ee
It is worth noting that the continuous equation (\ref{5.10}) can be a;so
obtained by taking the continuum limit of a suitable linear combination of
the $l=-1$, $l=0$, $l=1$ flows (eqs. (\ref{2.35b}), (\ref{2.36a}),
(\ref{2.38a}) ).

\section{Concluding remarks}
\label{sec6}
We would like to stress here that the system presented in this paper is, to
our Knowledge, the only integrable discrete version of AKNS hierarchy that
keeps the canonical Poisson structure of the continuum model and admits a
natural vector generalisation.
\par
As further developments of the research reported in this
paper, we mention the derivation of Backlund
tranformations for our lattice equations, to be considered as
integrable fully discrete two-dimensional lattices,
and the formulation of a proper quantum version in the framework of
the Quantum Inverse Method.
\par
Research has already started on both the above issues, with
encouraging preliminary results.

\clearpage

\newcommand{\Vol}[1]{{\bf #1}}

\end{document}